 \documentstyle[12pt,aasms]{article}

\newcommand{\bq}{\begin{equation}}
\newcommand{\eq}{\end{equation}}

\begin{document}
\def\refitem{\par\parskip 0pt\noindent\hangindent 20pt}

\title{The Velocity Field from Type Ia Supernovae Matches 
the Gravity Field from Galaxy Surveys}

\author{Adam G. Riess$^1$, Marc Davis$^1$, Jonathan Baker$^1$ \& Robert P. 
Kirshner$^2$}

\affil{$^1$Department of Astronomy, University of California, Berkeley, CA 
94720-3411 \\ $^2$Harvard-Smithsonian Center for Astrophysics, 60 Garden Street, 
Cambridge, MA 02138}
  
\begin{abstract}
We compare the peculiar velocities of nearby SNe Ia with those
predicted by the gravity fields of full sky galaxy catalogs.  The
method provides a powerful test of the gravitational instability
paradigm and strong constraints on the density parameter $\beta
\equiv \Omega^{0.6}/b$.  For 24 SNe Ia within 10,000 km s$^{-1}$,
we find the observed SNe Ia peculiar velocities are well modeled
by the predictions derived from the 1.2 Jy IRAS survey and the
Optical Redshift Survey (ORS).  Our best $\beta$ is 0.4 from
IRAS, and 0.3 from the ORS, with $\beta>0.7$ and $\beta<0.15$
ruled out at 95\% confidence levels from the IRAS comparison.
 Bootstrap resampling tests
show these results to be robust in the mean and in its error.
The precision of this technique will improve as additional nearby
SNe Ia are discovered and monitored.
 
\end{abstract}
subject headings:  supernovae:general ; cosmology:observations, large-scale
structure of universe---Local Group

\vfill
\eject
 
\section{Introduction}

The motions of galaxies probe the size of potentials formed by
the gravitating matter in the Universe.  Redshift surveys map the spatial
distribution of luminous matter.  Together, the two measure the
contribution of luminous and dark matter to the contents of the Universe,
providing a direct measure of the density parameter on the largest 
possible scale.

Previous attempts to test the gravitational instability paradigm and
constrain the mass density parameter, $\Omega_0$, are hindered by imprecise 
distance estimates to
 individual galaxies.
  Distance indicators based on empirical
relationships between galaxy luminosity and internal velocity (i.e.,
Tully-Fisher and $D_n-\sigma$) yield individual distance uncertainties of 
20-25\%
(Jacoby et al 1992, Tully \& Fisher 1977, Dressler
et al 1987, Willick et al 1996).  To
increase the utility of such data, workers have amassed it in great
quantity.  While this strategy diminishes the random component of error,
imprecise distances give rise to more troublesome systematic
errors including selection bias, Malmquist bias,
and smoothing biases (Strauss and Willick 1995).  The antidote for such
biases depends on knowledge of the distance uncertainty, a controversial 
quantity (Willick 1995,
Mathewson \& Ford 1994, Federspiel, Sandage, \& Tammann 1994, Giovanelli \&
Haynes 1996).  The forced marriage of inhomogeneous catalogs could result in
 additional errors in the inferred velocity
field.  Despite these challenges, remarkable progress has been made in this
 field 
(Dekel 1994,
Dekel, Bertschinger, \& Faber 1990, Davis, Nusser, \& Willick 1996, Hudson
1994, Shaya, Peebles, \& Tully 1995, Nusser \& Davis 1994).

Type Ia supernovae are well suited to provide an independent test of the
gravitational instability paradigm and to constrain the mass density.  Light
 curves from the Cal\'{a}n/Tololo Survey (Hamuy et al 1993, 1996)
and the CfA survey (Riess 1996, Riess et al 1997) yield distances with
5-10\% uncertainty over the redshift range  1000 $\leq cz \leq $36000 km 
s$^{-1}$. Although the sample of observed SNe Ia is relatively small,
 the depth and precison of supernova distances provide some advantages.  For
 the reduction of random errors, one
SN Ia is worth $\sim$ 10 Tully-Fisher or Dn-$\sigma$ measurements.
Systematic bias, which rises with the square of the distance uncertainty,
is also $\sim$ 10 times smaller for SNe Ia distances.  Individual distance
errors can be derived from SNe Ia light curves (Riess, Press, and Kirshner
1995a, 1996) which give meaningful measurements of the deviation
from smooth Hubble flow.  Combining the SNe Ia data with models of the
predicted ``peculiar velocities", one can constrain the 
 mass density traced by galaxy fluctuations while testing the
gravitational instability model for structure formation.

\section{Analysis}

The direct comparison of observed peculiar velocities and mass density
fluctuations  is not trivial.  The problem becomes
tractable only by assuming that the large scale
component of the flow is single-valued and irrotational.  Since galaxy 
fluctuations on scales larger than $10h^{-1}$ Mpc are
observed to be less than unity (Davis and Peebles 1983), 
these assumptions are reasonable. 
Furthermore, for such fluctuations at late times, 
linear perturbation theory is quite accurate, and the expected
velocity field ${\bf v_p}$ can be related 
to the peculiar gravity field ${\bf g}$ (Peebles 1980) by
\bq {\bf v_p(r)} = \left({2 \over 3\Omega_0^{0.4} H_0} \right) 
{\bf g(r)} \quad .\eq
This equation simply states that the observed peculiar velocity results from the 
gravitational acceleration acting over a Hubble time.

The gravity field ${\bf g(r)}$ may be inferred from the distribution of
galaxies which are assumed to trace the matter field.  It is common to
assume an unknown but linear bias exists between galaxy fluctuations
$\delta_G$ and matter fluctuations $\delta_\rho$, by employing a bias
parameter $b$, i.e.  $\delta_G = b \delta_\rho$.  This simplification,
while not accurate on small scales or in dense environments, should suffice
on large scales.
Using linear theory, one can constrain only a combination of
$\Omega_0$ and $b$, $\beta \equiv \Omega_0^{0.6}/b$.  A variety of methods
for estimating ${\bf g(r)}$ from galaxy surveys are reviewed by Strauss and
Willick (1995).  Here, we use the method of Nusser and Davis (1994),
which can successfully repair the gravitational distortions in redshift space.
However, this technique cannot properly treat regions with multivalued relations 
between velocity
and distance, expected near clusters of galaxies.  The minimum smoothing
scale of the derived gravity field is 500 km s$^{-1}$, and this smoothing
increases with distance in proportion to the mean interparticle spacing of
the galaxy catalog.  This increase is necessary to suppress artificial two-body
acceleration (Strauss et al 1992).

%

For comparison to the SNe Ia peculiar velocities, we computed the gravity
field from the 1.2 Jy flux limited IRAS redshift survey of galaxies (Fisher
et al 1995).  This catalog contains 6010 galaxies with median redshift of
6000 km s$^{-1}$ and provides a useful measure of the density field out to $cz
\approx 15,000$ km s$^{-1}$.  The IRAS catalog is known to underrepresent the elliptical-rich cores of
clusters relative to optically catalogs.  Thus it is 
of interest to use the ORS catalog of optically selected
galaxies for an alternative calculation of the gravity field (Santiago et al 1995, Baker
et al 1997).  The ORS contains 8457 galaxies, 
supplemented by the 1.2 Jy IRAS survey in sky regions without
optical coverage. 


We measured the distances and their uncertainties for 25 SNe Ia within
10,000 km s$^{-1}$ using MLCS distance measurements (Riess, Press, and
Kirshner 1996).  The redshifts of the SNe Ia come from Hamuy et al (1996),
Riess (1996), and Riess et al (1997).  Because of the limitation of linear
biasing and the risk of multivalued flows, we
conservatively discarded one SN Ia, SN 1992G, due to its proximity to the
Virgo Cluster.  In the directions of the remaining SNe, we computed the
expected distance-redshift relations based on the gravity fields of either
the ORS or IRAS surveys, as a function of $\beta$.  Figure 1 shows an
example of these curves. We assume a redshift error of 200 km s$^{-1}$ as an
estimate of the
small scale component of the radial peculiar velocity not describeable by
linear theory.

For each SN, we have the distance $d_i$ (in km s$^{-1}$), a distance error
$\delta d_i$ (in km s$^{-1}$), a redshift $z_i$, and a ``redshift error", 
$\delta
cz_i =200$ km s$^{-1}$.  From the gravity maps,  we have the
functions $z(j,\beta)$ and $d(j,\beta)$ 
along a set of points $j$ towards each SN's direction.  We seek the minimum 
separation of each point from the
predicted curves, in units of the standard deviations.  That is, for the
{\it ith} SN Ia, we compute a contribution to a $\chi^2$, 
\bq \delta
\chi(\beta)^2_i = {\rm min_j}\left( \left({d_i - d(j,\beta)\over \delta
d_i}\right)^2 + \left({z_i - z(j,\beta)\over \delta z_i}\right)^2 \right),
\eq 
where the minimization is over the locus of points $j$ which
defines the curve $z(d,\beta)$ toward each SN.
 A goodness
of fit is computed by summing over all the SNe, \bq \chi(\beta)^2 =
\sum_i \delta \chi(\beta)^2_i \quad , \eq with results shown at the top of 
Figure 2 for
both the IRAS and ORS surveys.

For the IRAS survey, we find 
$\beta_{IRAS} = 0.40 \pm 0.15$ and for the ORS survey we find
$\beta_{ORS}= 0.30 \pm 0.10$.  In both cases the value of $\chi^2$ at the
minimum is within the expected tolerance, confirming gravitational instability as a valid model for the observed peculiar motions of
SNe Ia.  We find ${\beta_{ORS} \over \beta_{IRAS}}=0.75 \pm
.38$, in good agreement with the relative biasing of 0.7
derived from the
correlation functions of ORS and IRAS galaxies (Fisher et al 1994).

The predicted and observed peculiar velocity fields for the best values
of $\beta$ are shown in Figure 3 in the Local Group (LG) frame at the location 
of each of the 24 SNe Ia; these numbers are also listed in Table 1.
Although the gravity predictions are derived
independently from the observed peculiar velocities, 
the similarity of the two is
remarkable.  
The leverage any
single SN Ia measurement has in determining $\beta$ depends on its
location.  Because the dominant feature of the flow
is dipolar, SNe Ia along the axis of this dipole carry more weight than
those whose radial motion is perpendicular to it.  Although the supernova data 
can be extended to greater depth, 
beyond 10,000 km
s$^{-1}$, the sampling of the IRAS and optical surveys is inadequate to
derive useful peculiar velocity predictions.


\section{Tests of Robustness}

To verify that our results are free from the vagaries of 
our supernova and galaxy
samples, we performed bootstrap resampling tests on both data sets.  This
procedure tests the effects our choice of sample points and their
uncertainties have on the estimates of $\beta$ by drawing new samples of
data from our best estimate of the underlying population: our sample (Press
et al. 1992).

We first drew 200 sets of randomly chosen SNe Ia from the sample of 24
SNe Ia, drawing each object with a Poisson probability of expectation value
of 1.  We then subjected each set to our maximum likelihood estimator
for $\beta$, using the gravity field
of the IRAS catalog.  
As seen on the bottom of Figure 2, the distribution
for $\beta$ preferred by the individual SNe Ia agrees well with the single
likelihood distribution estimated from our sample. 

Similarly, we  tested the robustness of the IRAS gravity field
by generating 20 bootstrap resampled IRAS catalogs, as we did for the
resampled SNe catalogs.  For each resampled IRAS catalog, we generated
 full gravity fields for the range $0.1 \le \beta \le 1.2$.  The resulting
$\beta$ values derived from the minimum $\chi^2$ of the 24 SNe sample
are also shown at the bottom of Figure 2.    Again, this distribution is
 consistent with the constraints for
$\beta$ inferred from the $\chi^2$ of the original datasets.  Thus we believe
the estimates in \S 2 are reasonable.



\section{Discussion}











	Figure 3 demonstrates the remarkable consistency between the observed
 peculiar velocities of 24 SNe Ia and those predicted from the gravity fields
 of optical or IRAS galaxies using linear perturbation theory and the best value 
for $\beta$.  The excellent $\chi^2$ fits in Figure 2 confirm our simple model 
for the source of peculiar velocities while putting useful constraints on the 
mass density parameter, $\beta$.  

     The signals, seen in Figure 3, are largely dipole patterns revealing
the motion of the LG relative to the SNe frame.  They constrain the {\it 
shear} in the large scale velocity field induced by the gravity of galaxies 
within the sample's 10,000 km s$^{-1}$ radius.  The bulk of
the CMB dipole signature does appear to have been generated
within this radius.  Velocity dipole patterns that match the gravitational
dipole signature have been detected in several other surveys (e.g.
Riess, Press, and Kirshner 1995b, da Costa et al 1996; Giovanelli et al 
1996), but flows inconsistent with
the predicted gravity field for all $\beta$ values
have also been reported (Lauer and Postman,
1992, 1994; Davis, Nusser, and Willlick 1996). 

	A more sophisticated, normal
mode comparison of the IRAS gravity field to the velocity field
derived from a sample of 2900 Mark-III Tully-Fisher galaxies
within a limiting redshift of 6000 km s$^{-1}$ shows
inconsistencies for any value of $\beta$ (Davis, Nusser, and
Willick 1996).  This same procedure applied to the SFI catalog yields consistent gravity and velocity fields with $\beta_{IRAS}=0.6\pm0.1$.
The POTENT reconstruction of the local density field from the
Mark-III catalog recovers many observed features in the IRAS maps
and estimates $\beta =0.89 \pm 0.12$, but some inconsistencies
persist (Sigad et al 1997, Dekel 1994).  The failure to match the
fields remains unexplained.

	By limiting the analysis to a redshift of 3000 km s$^{-1}$, Willick
et al (1996) successfully
 applied the VELMOD algorithm to compare the IRAS gravity field to
a sample of 838 Tully-Fisher galaxies; a maximum likelihood analysis leads to
$\beta =0.49 \pm 0.07$, is consistent with our results
for the IRAS gravity field.    Similar low values
of the density parameter emerge from the least action method applied to the flow
field with 3000 km s$^{-1}$ (Shaya, Peebles, and Tully 1995).  
  These procedures are distinct
and do not all suffer from the same biases.  Thus it is encouraging that they 
are 
leading to consistent (and perhaps reliable) results.

In Figure 3, the individual peculiar velocities in the SN map
appear to be slightly larger than the velocities predicted at the
same locations by the IRAS or ORS fields.  We attribute this to
measurement noise and velocity noise (small scale velocity flows)
present in the SN data but not in the heavily smoothed IRAS or
ORS velocity fields.  Because of this noise, some SN velocity residuals 
do not match the sign (filling
in Figure 3) of the IRAS predictions and/or the mode of the signs of the
hemisphere in which they occur.  But overall, the $\chi^2$
minimization for the supernovae takes these sources of error
properly into account and produces the best matched $\beta$ for
the data sets.  Our determinatiuons of $\beta$ are not sensitive to our estimate of the velocity noise or SNe Ia errors, the latter limited by the small dispersion from smooth Hubble flow.  As the
distances for individual supernovae increase, the errors (in
km s$^{-1}$) increase, while for the IRAS and ORS data sets the
smoothing becomes stronger and the amplitude of point to point
variations gets smaller.  Although more distant SNe carry less weight, we find the objects within as well as beyond 5000 km s$^{-1}$ give consistently low values for $\beta$.  Nusser and Davis (1995) show how to
avoid the effects of mismatched smoothing that appear here, but
the present SN sample is too sparse to apply their methods.

     MLCS distances are precise enough to characterize the peculiar velocity 
field in the direction of each supernova.   Yet, during this application we 
found instrinsic uncertainties still limit the precision of relative SNe Ia 
distances to no better than 5\%. Future investigations into SNe observables
 may improve the precision of these distances or our understanding of their 
limitations.

  As the supernova data set grows, the precision
of this comparison between gravity and velocity will improve, and
the methods of analysis that have been used on the galaxy data
sets will become appropriate.  No tool for mapping the peculiar velocity field 
has a brighter future.

{\bf Acknowledgements}

This work was supported by NSF grant AST95-28340 and NASA grant NAG 5-1360 at 
UCB, NSF grants AST95-28899 and AST96-17058 at Harvard University and by the 
Miller Institute for Basic Research in Science through a fellowship to A.G.R..





\begin{table}[bp]    
\begin{center}    
\begin{tabular}{lccccccc} 
\multicolumn{8}{c}{Table 1: Peculiar Velocity Data} \\
\hline \hline
  &  &  &  $cz$ &  $cz-H_0d$ & $\sigma_d$ & IRAS $v$ & ORS $v$ \\
  {\em SN Ia}     &   $ l $ & $ b $ &  (km s$^{-1}$)  &  (km s$^{-1}$) & (km 
s$^{-1}$) & $(\beta=0.4)$ & $(\beta=0.3)$ \\

(1) & (2) & (3) & (4) & (5) & (6) & (7) & (8) \\
\hline   
1995al&  192.60&   51.40&    1493.&   -466.& 123.&	-407.& -365.\\
1996X&   310.20&   35.70&    1845.&   -96.& 149.&	-81.&   -204.\\
1995D&  230.00&   39.67&     2000.&   -258.&	156.&	 -457.& -191.\\
1996Z&  253.60&   22.60&     2014.&   -427.& 289.&	-320.&  -71.\\
1991M&   30.39&   45.90&     2489.&   15.& 201.&	-183.&   -229.\\
1992K&  306.28&   16.31&     2825.&   -381.& 345.&	-18.&  17.\\
1995E&  141.97&   30.27&     3639.&   175.&  217.&	21.&   -368.\\
1991ag&  342.56&  -31.64&    4150.&   -131.& 289.&	176.&  358.\\
1992al&  347.30&  -38.50&     4355.&  566.&  245.&	 187.& 373.\\
1994S&  187.84&   85.75&     4539.&   81.& 283.&	-129.&   -167.\\
1995bd&  187.10&  -21.70&     4808.&  227.&  329.&	 46.&  193.\\
1993ae&  144.62&  -63.23&     5521.&  333.&  409.&	226.&  139.\\
1992bo&  261.88&  -80.35&     5662.&  424.&  369.&	 99.&  327.\\
1992bc&  245.70&  -59.64&     6053.&  551.&  355.&	-59.&  282.\\
1994M&  291.69&   63.03&     6730.&   -716.& 564.&	-407.& -29.\\
1995ak&  169.70&  -49.00&     6887.&  1063.& 578.&	 252.& 221.\\
1993H&  318.20&   30.30&     6982.&   263.& 443.&	-156.&  62.\\
1992ag&  312.50&   38.40&     7295.&  -489.& 984.&	 -272.&	78.\\
1992P&  295.62&   73.11&     7447.&   -519.& 539.&	-408.& -108.\\
1994Q&   99.60&   65.00&     8956.&   -790.&  648.&	-12.& -632.\\
1996C&   64.38&   39.68&     8872.&   -535.&  819.&	137.& -477.\\
1993ah&   25.90&  -76.80&     8974.&   -53.&  1012.&	183.& 277.\\
1990O&   37.60&   28.40&     9247.&    -1062.& 749.&  5.&	-197.\\
1991U&  311.82&   36.21&     9290.&   180.& 1357.&	-260.&  23.\\
\hline \end{tabular}
\end{center}
\end{table}

\vfill \eject

\centerline {\bf References}
\vskip 12 pt

\refitem Baker, J., Davis, M., Strauss, M., Lahav, O., Santiago, B. X., 1997, 
in preparation

\refitem da Costa, L., Freudling, W., Wegner, G., Giovanelli, R., 
Haynes, M., \& Salzer, J. 1996,Ap.J.L., 468, L5

\refitem da Costa, L., et al, 1997, in preparation

\refitem Davis, M. \& Peebles, 1983, ApJ, 267, 465

\refitem Davis, M., Nusser, A., and Willick, J. 1996, ApJ, 473, 22

\refitem Dekel, A., 1994, ARA\&A, 32, 371

\refitem Dekel, A., Bertschinger, E. \& Faber, S. M., 1990, ApJ, 364, 349

\refitem Dressler, A., Faber, S. M., Burstein, D., Davies, R. L., Lynden-Bell, 
D., Terlevich, R. J. \& Wegner, G. 1987, ApJ, 313, L37

\refitem Federspiel, M., Sandage, A. \& Tamman, G. A., 1994, ApJ, 430, 29

\refitem Fisher, K., Huchra, J. P., Strauss, M. A., Davis, M., Yahil, A., 
Schlegel, D., 1995, ApJS, 100, 69

\refitem Fisher, K., Davis, M., Strauss, M. A., Yahil, A., Huchra, J. P., 1994, 
MNRAS, 266, 50

\refitem Giovanelli, R., Haynes, M., Wegner, G., da Costa, L., Freudling, W.,
\& Salzer, J. 1996, ApJ, 464, L99 

\refitem Hamuy, M., Phillips, M. M., Suntzeff, N.B., Schommer, R. A., Maza, 
R.A., \& Aviles, R. 1997, AJ, 112, 2048

\refitem Hamuy, M., et al 1993, AJ, 106, 2392

\refitem Hudson, M. J., 1994, MNRAS, 266, 468

\refitem Jacoby, G., et al, 1992, PASP, 104, 599

\refitem Lauer, T., \& Postman, M. 1992, Ap.J.Lett., 400, L47

\refitem Lauer, T., \& Postman, M. 1994, Ap.J. 425, 418

\refitem Mathewson, D. S. \& Ford, V. L., 1994, ApJ, 434, L39

\refitem Nusser, A. \& Davis, M., 1994, ApJ, 421, L1

\refitem Nusser, A. \& Davis, M., 1995, MNRAS, 276, 1391

\refitem Peebles, P.J.E. 1980, The Large Scale Structure of the Universe, 
Princeton
U. Press.

\refitem  Press, W.H., Teukolsky, S.A., Vetterling, W.T. \& Flannery, B.P. 1992, 
{\it Numerical Recipes, 2nd ed.} (Cambridge University Press)

\refitem Riess, A.G., Press W.H., Kirshner, R.P., 1995a, ApJ, 438 L17

\refitem Riess, A.G., Press W.H., Kirshner, R.P., 1995b, ApJ, 445, L91

\refitem Riess, A.G., Press, W.H., \& Kirshner,  R.P. 1996, ApJ, 473, 88

\refitem Riess, A.G., PhD thesis, 1996, Harvard University

\refitem Riess, A. G., et al 1997, in preparation

\refitem Santiago, B. X., et al, 1995, ApJ, 446, 457

\refitem Sigad, Y., Dekel, A., Strauss, M., \& Yahil, A. 1997, in preparation

\refitem Shaya, E., Peebles, P., and Tully, B. 1995, Astrophysical Journal, 454,
15.

\refitem Strauss, M., \& Willick, J. 1995, Physics Reports, 261, 271

\refitem Strauss, M. A., Yahil, A., Davis, M., Huchra, J. P.\& Fisher, K. B., 
1992, ApJ, 397, 395

\refitem Tully, R. B. \& Fisher, J. R., 1977, A\&A, 54, 661

\refitem Willick, J., Strauss, M., Dekel, A., \& Kolatt, T. 1996, preprint 
astro-ph/9612240

\refitem Willick, J. A., Courteau, Faber, S. M., Burstein, D. \& Dekel, A., 
1995, ApJ, 446, 12

\vfill \eject

\begin{figure}
\vspace*{130mm}
\includegraphics{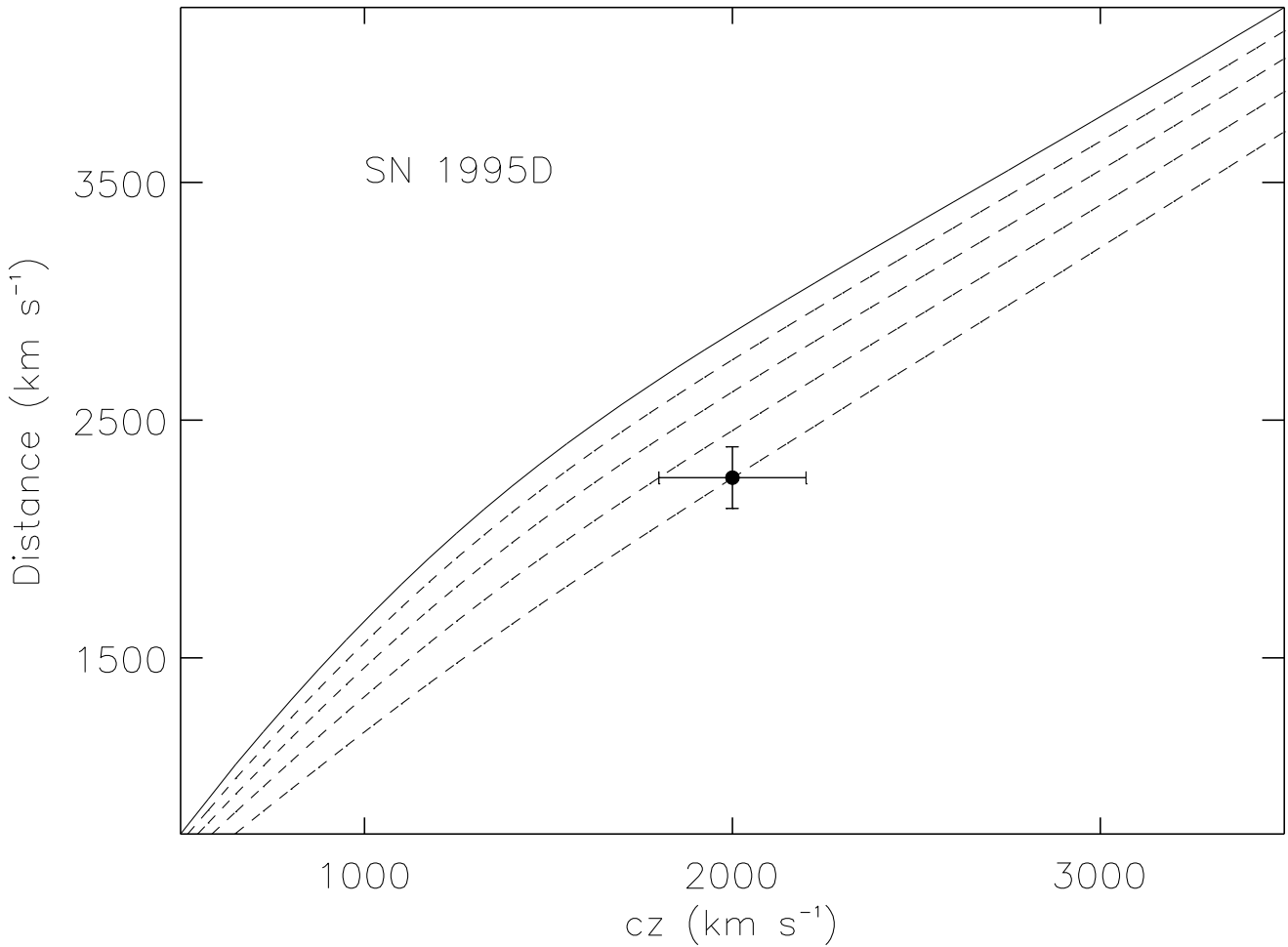}
\caption {Velocity-distance relation in the vicinity of SNe Ia.  This
 is an example of the velocity versus distance predictions of the IRAS 
gravity maps as a function of the mass density parameter, $\beta$, in the 
direction of SNe Ia.  The solid line shows the $\beta=1.0$ prediction, with 
subsequent dotted lines showing the predictions for $\beta$ in decreasing 
increments of 0.2.  Overplotted is the measurement of the SNe Ia redshift, 
MLCS distance (in km s$^{-1}$) and distance error with a small scale velocity 
error of 200 km s$^{-1}$.  Such plots contributes to the determination of $\beta$ via equation (2).}
\end{figure}
\vfill
\eject
\begin{figure}
\vspace*{170mm}
\includegraphics{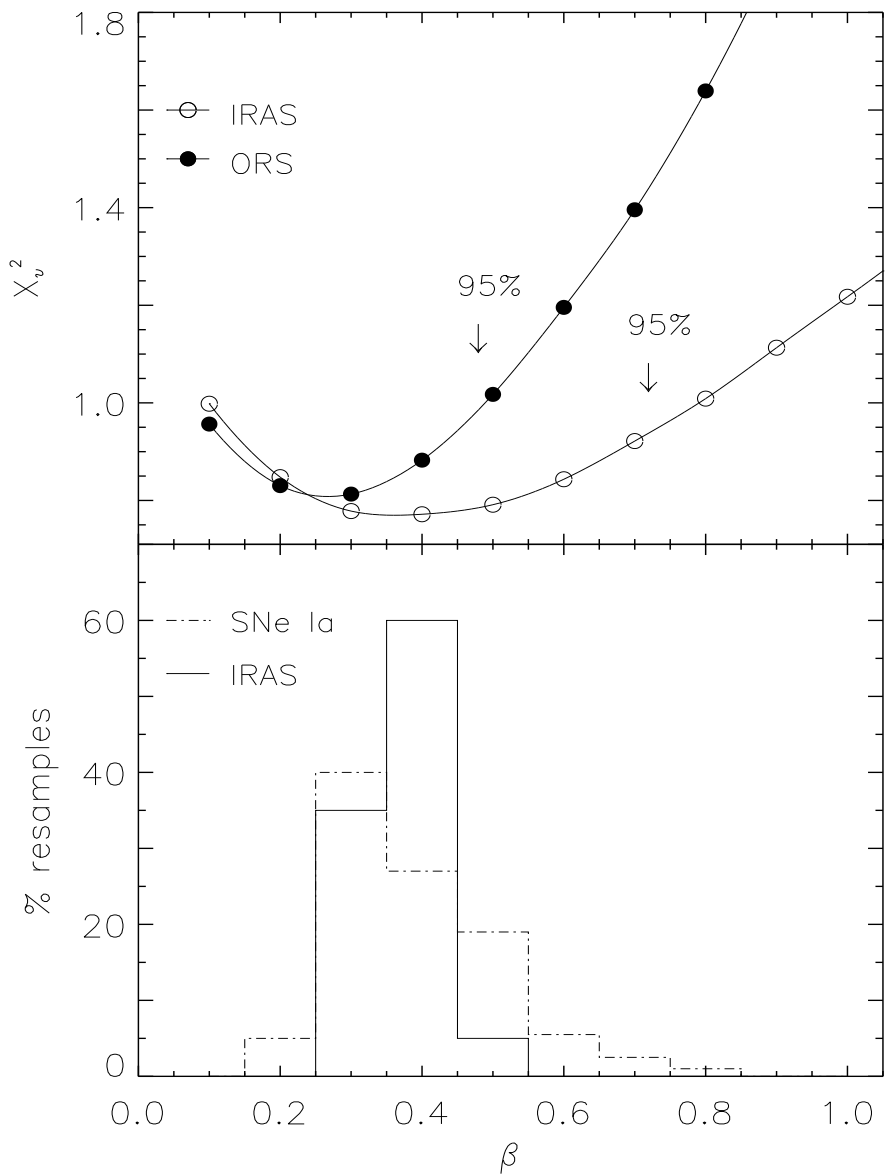}
\caption {Constraints on the mass density parameter,
$\beta$, and tests of robustness. The top panel shows the value
of our $\chi^2$ statistic as a function of $\beta$, equation (3),
reduced (divided) by the twenty-three degrees of freedom.  Our
best $\beta$ is 0.4 from IRAS, and 0.3 from the ORS, with
$\beta>0.7$ and $\beta<0.15$ ruled out at 95\% confidence levels
for the comparison to IRAS.
Bootstrap resamplings (bottom panel) of the SNe Ia and IRAS
galaxies, descibed in \S 3, validate our estimates of $\beta$.}
\end{figure}
\vfill
\eject
\begin{figure}
\vspace*{170mm}
\includegraphics{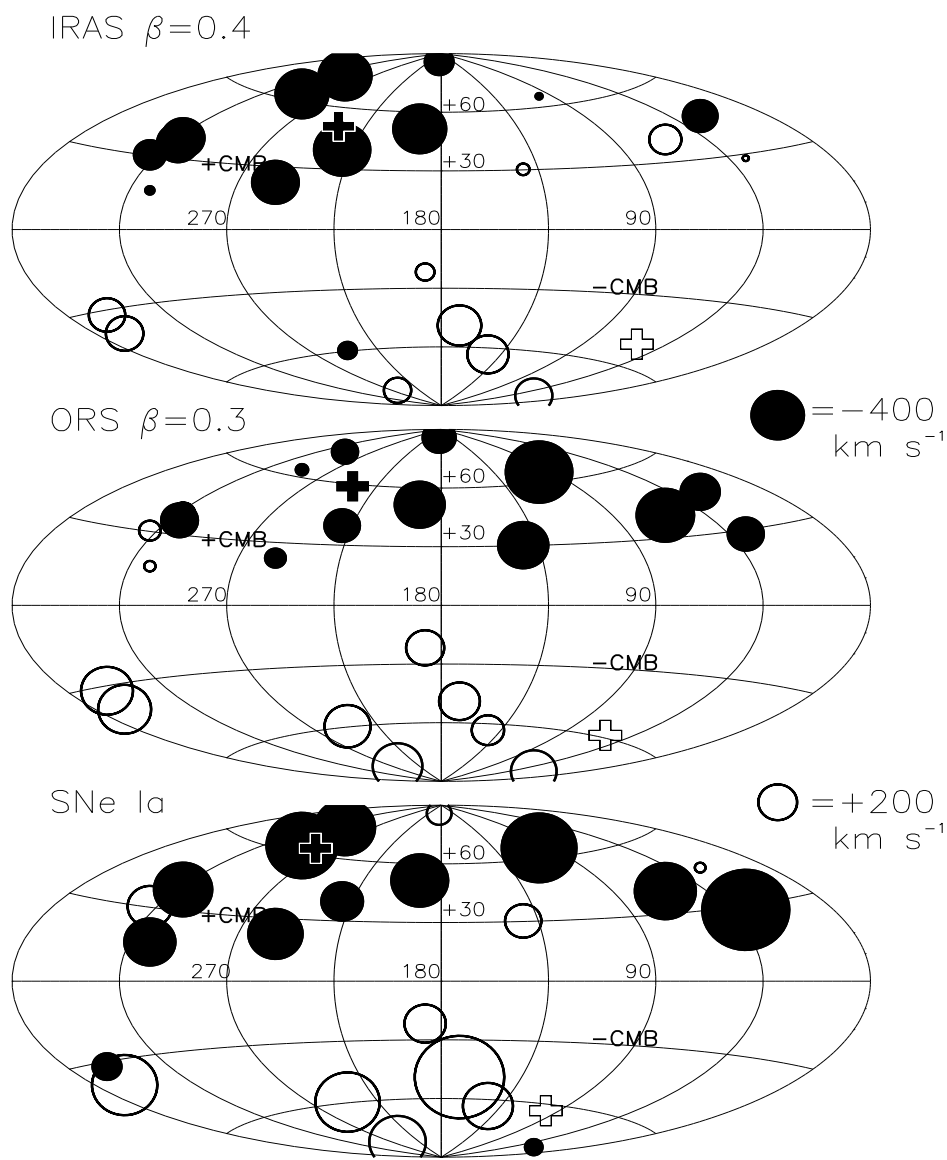} 
\caption {Predicted and observed peculiar velocity fields in the Local 
Group rest frame.  Filled/Open points represent SNe Ia with measured 
negative/positive peculiar velocities (bottom map). The top two maps show the 
peculiar velocities predicted by the gravity fields of the IRAS and ORS catalogs at the position of the SNe Ia for values of $\beta$ which best fit the SNe Ia velocity field.  Filled/Open crosses mark the direction towards which the Local Group is approaching/receding. The directions of increased/decreased cosmic 
microwave background temperature are indicated by +CMB/-CMB.}
\end{figure}
\vfill
\eject

\end{document}